\documentclass[vecphys]{svmult}

\usepackage{makeidx}         
\usepackage{graphicx}        
\usepackage{multicol}        
\usepackage[bottom]{footmisc}

\makeindex             

\begin{document}

\title*{Discovery of microquasar LS I +61 303 at VHE gamma-rays with MAGIC}
\author{N\'uria Sidro\inst{1}
for the MAGIC Collaboration\inst{2}}
\institute{Institut de F\'isica d'Altes Energies (IFAE), Universitat Aut\`onoma de Barcelona, 08193, Bellaterra, Spain
\texttt {nuria.sidro@ifae.es}
\and Updated collaborator list at \texttt{http://wwwmagic.mppmu.mpg.de/collaboration/members}}
%
%
\maketitle
\index{N. Sidro}

\abstract{
Here we report the discovery of very high energy ($>100$GeV) $\gamma$-ray
emission from the radio emitting X-ray binary LS I +61 303 with the
Major Atmospheric Gamma Imaging Cherenkov (MAGIC) telescope. 
This high energy emission has been found to be variable, detected 4
days after the periastron passage and lasting for several days. 
The data have been taken along different orbital cycles, and the fact
that the detections occur at similar orbital phases, suggests that the
emission is periodic. 
Two different scenarios have been involved to explain this high energy
emission: the microquasar scenario where the $\gamma$-rays are produced
in a radio-emitting jet; or the pulsar binary scenario, where 
they are produced in the shock which is generated by the interaction
of a pulsar wind and the wind of the massive companion.
}
\section{Introduction: The MAGIC Telescope} 
\label{sec:intro}

The Major Atmospheric Gamma Imaging Cherenkov 
(MAGIC) telescope~\cite{performance} is a very high energy (VHE) $\gamma$-ray
telescope, operating in a energy band from 100 GeV to 10 TeV,
exploiting the Imaging Air Cherenkov (IAC) technique. Located on the
Canary Island of La Palma, at $28^\circ 45^\prime 30^{\prime\prime}$N,
$17^\circ$ $52^\prime$ $48^{\prime\prime}$W and 2250~m above sea level. 
The telescope has a 17-m diameter tessellated parabolic mirror, and is
equipped with a 3.5$^\circ$-3.8$^\circ$ field of view
camera. See~\cite{martinez} for a complete description of the
instrument.

\section{The $\gamma$-ray binary LS~I~+61~303} 
\label{sec:lsi}

LS~I~+61~303 belongs, together with LS~5039~\cite{hessls} and
PSR~B1259-63~\cite{1259}, to a new class of objects, the so-called
$\gamma$-ray binary systems, whose electromagnetic emission extends up
to the TeV domain. 
In section~\ref{sec:scenarios} we will describe the possible VHE emission
scenarios for this object: it was proposed as microquasar candidates,
but also thought to be in a pulsar binary scenario.
Thus, the question of whether the three known $\gamma$-ray binaries
produce TeV emission by the same mechanism, and by which one, is
currently object of an intense debate~\cite{perspectives}.

\subsection{Description of the object}

This $\gamma$-ray binary system is composed of a B0 main sequence star
with a circumstellar disc, i.e. a Be star, located at a distance of
$\sim$2 kpc. A compact object of unknown nature (neutron star or black
hole) is orbiting around it, in a highly eccentric ($e=0.72\pm0.15$)
orbit.  

The orbital period --with associated
radio~\cite{gregoryold} and X-ray~\cite{taylor} outbursts-- is 26.496
days and periastron passage is at phase $0.23\pm0.02$~\cite{casares}.  

Radio outbursts are observed every orbital cycle at
phases varying between 0.45 and 0.95 with a 4.6 years
modulation~\cite{gregorynew}.

High-resolution radio imaging techniques have
shown extended, radio-emitting structures with angular extension of
$\sim$0.01 to $\sim$0.1 arc-sec, interpreted within the framework of
the microquasar scenario, where the radio emission is originated in a
two-sided, possibly precessing, relativistic jet
($\beta/c=0.6$)~\cite{massi}. 
However, the two-sided jets were not completely resolved and no
solid evidence of the presence of an accretion disk (i.e. a thermal
X-ray component) has been observed.  There are hints
of variability of the $\gamma$-ray flux~\cite{tavani}. 

LS~I~+61~303 was considered as one of the two microquasar candidates
positionally coincident with EGRET $\gamma$-ray
sources~\cite{kniffen}, and the only one located in the Northern
Hemisphere --hence a suitable target for MAGIC. 
The large uncertainty of the position of the EGRET source did not
allow an unambiguous association.

\subsection{MAGIC observations}

LS~I~+61~303 was observed with MAGIC for 54 hours
(after standard quality selection, discarding bad weather data)
between October 2005 and March 2006~\cite{lsi}. As mentioned above,
MAGIC is able to operate in moderate moon conditions, and in
particular, 22\% of the data used in this analysis were recorded under
moonlight.  The data analysis was carried out using the standard MAGIC
reconstruction and analysis software~\cite{magic_hess1813,magic_hess1834,magic_gc}.

\begin{figure}[hpt]
\centering
\includegraphics[width=10.0cm]{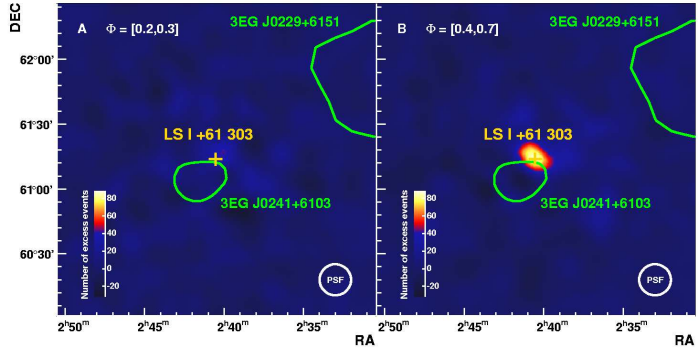}
\caption{
  Smoothed maps of $\gamma$-ray excess events above 400 GeV around
  LS~I~+61~303. (A) Observations over 15.5 hours corresponding to data
  around periastron (phase 0.2-0.3). (B) Observations over 10.7 hours
  at orbital phase 0.4-0.7. The position of the optical source
  LSI~+61~303 (yellow cross) and the 95$\%$ confidence level contours
  for two EGRET sources are shown. From Albert \emph{et
  al.}~\cite{lsi}.
}
\label{fig:lsi-skymap}
\end{figure}

\begin{figure}
\vspace{0.5cm}
\centering
\includegraphics[width=8.0cm]{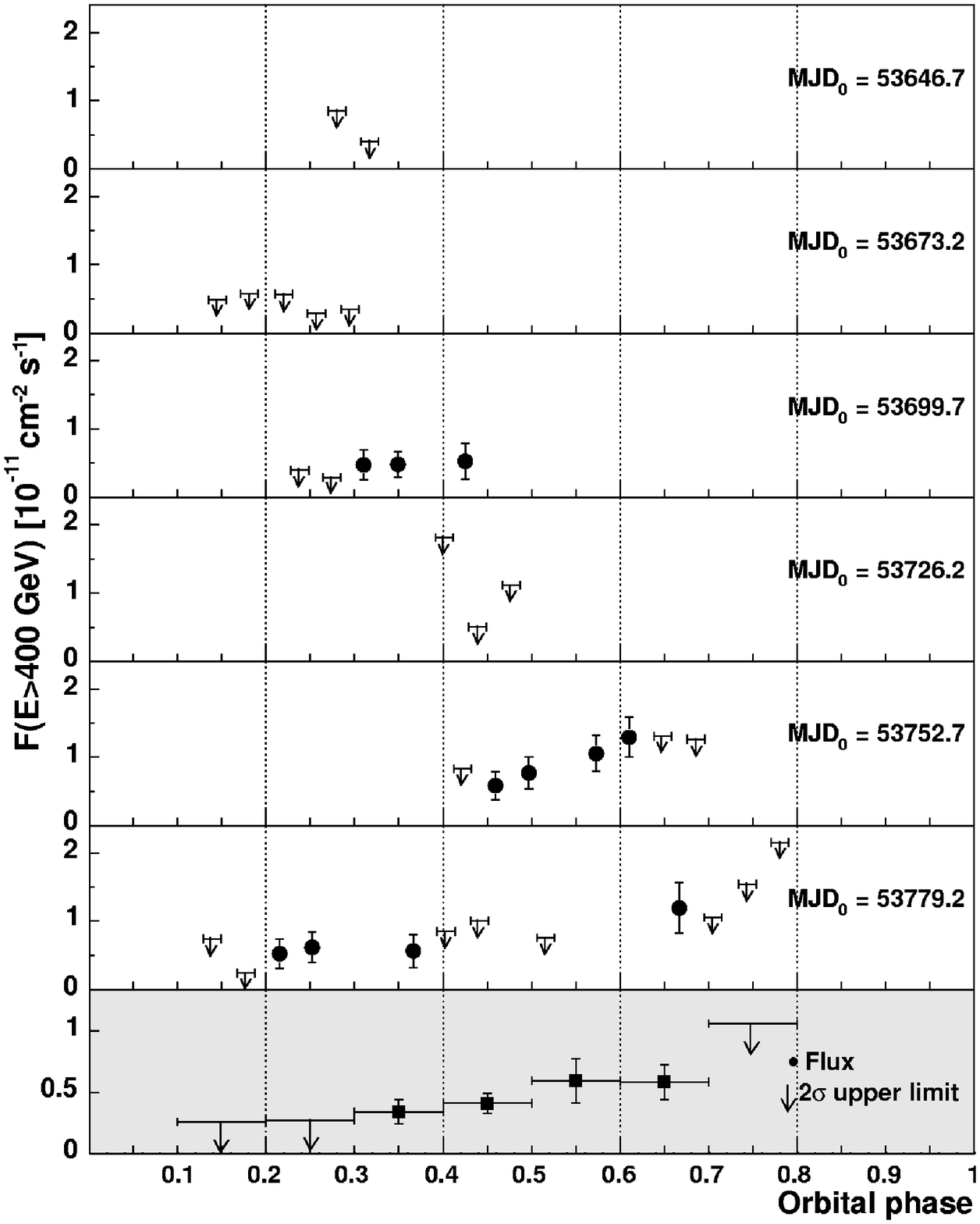}
\caption{VHE $\gamma$-ray flux of LS~I~+61~303 as a function of orbital
phase for the six observed orbital cycles (six upper panels, one point
per observation night) and averaged for the entire observation time
(bottom panel). Vertical error bars include 1$\sigma$ statistical error
and 10$\%$ systematic uncertainty on day-to-day relative fluxes. Only
data points with more than 2$\sigma$ significance are shown, and
2$\sigma$ upper limits~\cite{rolke} are derived for the rest. The modified
Julian date (MJD) corresponding to orbital phase 0 is indicated for
every orbital cycle. From Albert \emph{et al.}~\cite{lsi}.} 
\label{fig:lsi-lc}
\end{figure}

The reconstructed $\gamma$-ray map is shown in
Figure~\ref{fig:lsi-skymap}. The data were first divided into 
two different samples, around periastron passage (0.2-0.3)
and at higher (0.4-0.7) orbital phases. No significant excess in the
number of $\gamma$-ray events is detected around periastron passage,
whereas there is a clear detection (9.4$\sigma$ statistical significance) at
later orbital phases.

The distribution of $\gamma$-ray excess is
consistent with a point-like source located at (J2000):
$\alpha = 2^\mathrm{h}40^\mathrm{m}34^\mathrm{s}$, $\delta = 61^\circ
15^\prime 25^{\prime\prime}$, with statistical and systematic
uncertainties of $\pm 0.4^\prime$ and $\pm2^\prime$,
respectively. This position is in
agreement with the position of LS~I~+61~303. In the natural case in
which the VHE emission is produced by the same object detected at
EGRET energies, this result identifies a $\gamma$-ray source that
resisted classification during the last three decades.

Our measurements show that the VHE $\gamma$-ray emission from
LS~I~+61~303 is variable. The $\gamma$-ray flux above 400 GeV coming
from the direction of LS~I~+61~303 (see Figure~\ref{fig:lsi-lc}) has a
maximum corresponding to about 16$\%$ of the Crab nebula flux, and is
detected around phase 0.6. The combined statistical significance of
the 3 highest flux measurements is 8.7$\sigma$, for an integrated
observation time of 4.2 hours. The probability for the distribution of
measured fluxes to be a statistical fluctuation of a constant flux
(obtained from a $\chi^2$ fit of a constant function to the entire
data sample) is $3\times 10^{-5}$. The fact that the detections occur
at similar orbital phases hints at a periodic nature of the VHE
$\gamma$-ray emission. 

\begin{figure}
  \centering
  \includegraphics[width=6.7cm]{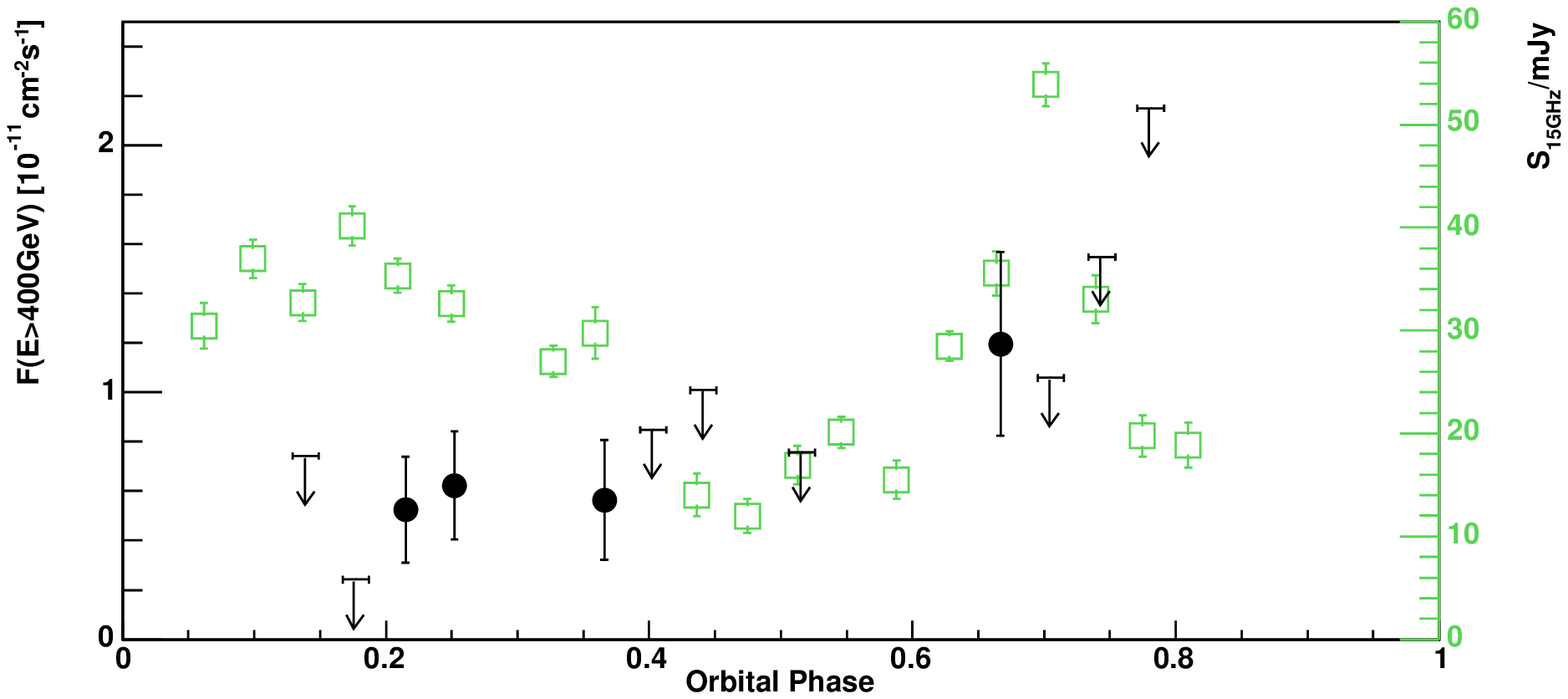}
  \includegraphics[width=6.1cm]{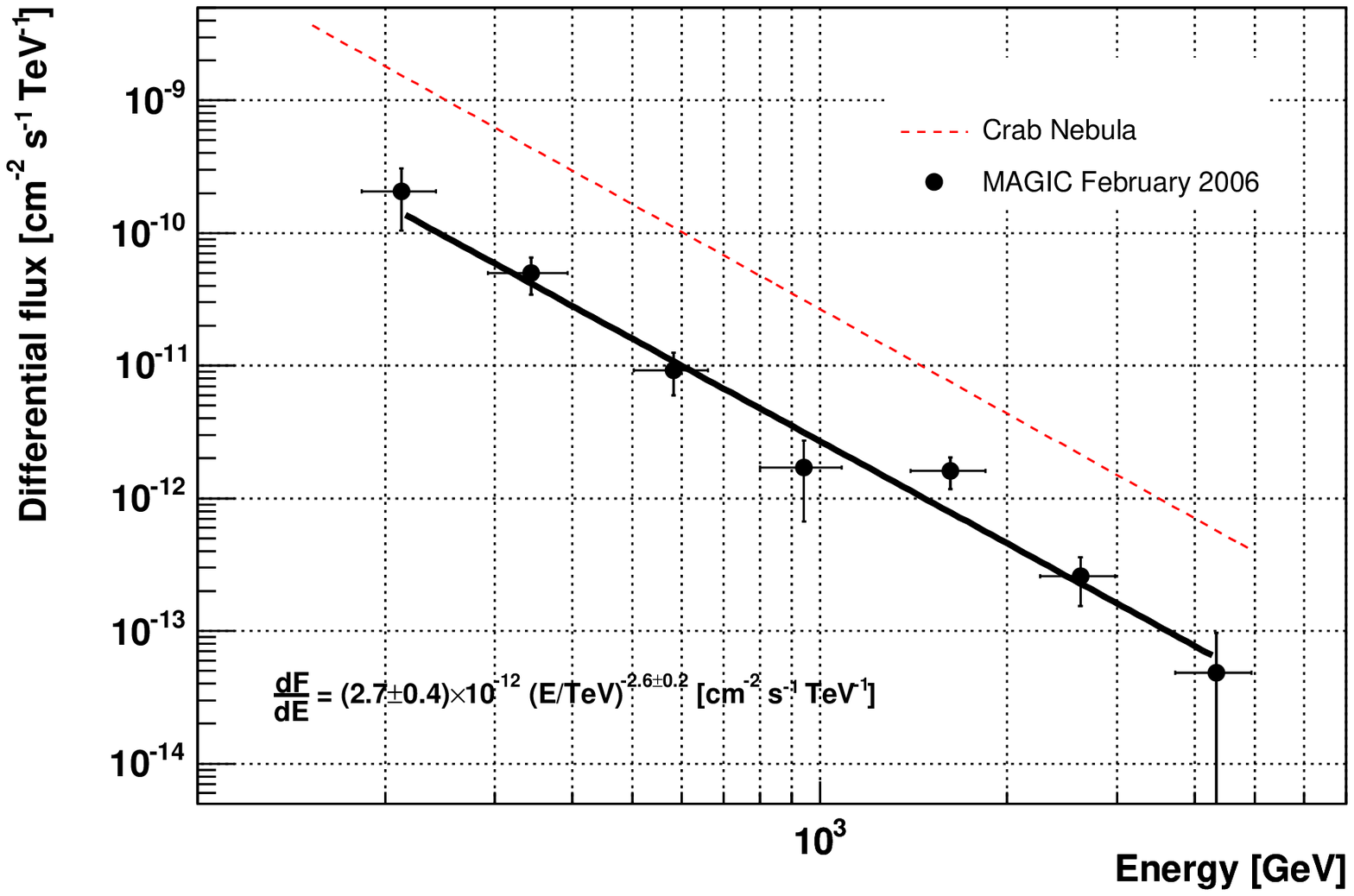}
  \caption{Left: LS~I~+61~303 radio flux density at 15~GHz measured with the
    Ryle Telescope (green squares, right axis) and results from the last
    orbital cycle observed by MAGIC (black dots, left axis). Right:
    Differential energy spectrum of LS~I~+61~303 for energies between 200
    GeV and 4 TeV and averaged for orbital phases between 0.4 and 0.7,
    measured by MAGIC. The dashed line corresponds to the Crab nebula
    differential spectrum also measured by MAGIC. The solid line is a
    fit of a power law to the measured points.
  } 
\label{fig:lsi-ryle-spec}
\end{figure}

Contemporaneous radio observations of
LS~I~+61~303 were carried out at 15 GHz with the Ryle Telescope
covering several orbital periods of the source. The peak of the radio
outbursts was at phase 0.7, i.e. between 1 and 3 days after the
increase observed at VHE $\gamma$-rays flux (see Figure~\ref{fig:lsi-ryle-spec}).

The VHE spectrum derived from data between $\sim$200 GeV and
$\sim$4~TeV at orbital phases between 0.4 and 0.7 is shown in
Figure~\ref{fig:lsi-ryle-spec}. The obtained spectrum is fitted
well ($\chi^2/ndf = 6.6/5$) by a power law function:
\begin{eqnarray*}
dN_{\gamma}/(dA/dt/dE) = (2.7 \pm 0.4 \pm 0.8) \times 10^{-12} \times E^{(-2.6 \pm 0.2 \pm 0.2)} \quad  \mathrm{cm}^{-2} \mathrm{s}^{-1} \mathrm{TeV}^{-1}
\end{eqnarray*}
%
where $N_{\gamma}$ is the number of $\gamma$-rays reaching the Earth per unit area
$A$, time $t$ and energy $E$ (expressed in TeV). Errors quoted are
statistical and systematic, respectively. This spectrum is consistent
with that measured by EGRET for a spectral break between 10 and
100~GeV. 

We estimate that the flux from LS~I~+61~303 above 200~GeV corresponds to an
isotropic luminosity of $\sim 7 \times 10^{33}$~erg~s$^{-1}$ (assuming
a distance to the system of 2~kpc~\cite{frail}). The intrinsic
luminosity is of the same order of that of the similar object
LS~5039~\cite{hessls}, and a factor $\sim 2$ lower than the previous
experimental upper limit ($< 8.8 \times 10^{-12}$~cm$^{-2}$ s$^{-1}$
above 500~GeV) obtained by Whipple~\cite{wipple}. 
LS~I~+61~303 displays more luminosity at GeV than at X-ray energies, a
behavior shared also by LS~5039.

\subsection{Emission scenarios}
\label{sec:scenarios}


Microquasars are a subclass of stellar, X-ray binary systems that
display prominent radio emission, usually attributed to the existence
of jets of relativistic particles.  They are named after the
similarities with active galactic nuclei (AGNs), since microquasars show
the same three ingredients that make up radio-loud AGNs: a compact
object, an accretion disc, and relativistic
jets~\cite{mirabel}. Hence, microquasars are galactic, scaled-down
versions of an AGN, where instead of a super-massive black hole we
deal with a compact object of just a few solar masses that accretes
material from a donor star.  The similarities with AGNs explain the
large interest risen by microquasars. 
In fact, the short timescale variability displayed by microquasars allows to see
changes in the ongoing physical processes within typical time scales
ranging from minutes to months, in contrast with the usual scales of
years to observe such variability in AGNs. In addition, microquasars
could measurably contribute to the density of galactic cosmic
rays~\cite{heinz}.

LS~I~+61~303 was usually thought to be similar to the microquasar LS~5039,
because evidence of relativistic jets has been found at
radio frequencies. In this scenario, the high energy emission would be 
produced in shocks at the relativistic jets~\cite{mirabel}. 
The observation of jets has triggered the study of different
microquasar-based $\gamma$-ray emission models, some regarding
hadronic mechanisms: relativistic protons in the jet interact with
non-relativistic stellar wind ions, producing $\gamma$-rays via
neutral pion decay; some regarding leptonic mechanisms:
IC scattering of relativistic electrons in the jet on stellar and/or
synchrotron photons~\cite{valenti}.

However, no clear signal of the presence of an
accretion disk (in particular a spectral feature between $\sim10$
and $\sim100$ keV due to the cut-off of the thermal emission) has
been observed so far. Because of that, it has been alternatively
proposed that relativistic particles could be injected into the
surrounding medium at the shock where the wind of the young pulsar and
the wind of the stellar companion collide~\cite{maraschi},
which seems to be the case of PSR~B1259-63. 

In the case of LS~I~+61~303 the resemblance of the time variability
and the radio/X-ray spectra with those of young pulsars support such
hypothesis. However, no pulsed emission has been detected from
LS~I~+61~303.  


Observation at VHE with MAGIC
simultaneously together with other instruments at other wavelength
domains --in particular radio-- will help elucidate the mechanism of
the TeV emission and hence the nature of $\gamma$-ray binaries.

\paragraph{Acknowledgements.}
We would like to thanks G. Pooley at Cavendish Laboratory (Cambridge,
UK) for providing us with the radio Ryle Telescope data. 
We also thank the IAC for the excellent working conditions at the
Roque de los Muchachos Observatory in La Palma. The support of the
German BMBF and MPG, the Italian INFN, the Spanish CICYT is gratefully
acknowledged. This work was also supported by ETH research grant
TH-34/04-3, and the Polish MNiI grant 1P03D01028.

%
%
%
%

\bibliography{biblio}

\begin{thebibliography}{10}

\bibitem{performance}
J.~Cortina et~al.
\newblock Technical performance of the {MAGIC} telescope.
\newblock Prepared for 29th International Cosmic Ray Conference (ICRC 2005),
  Pune, India.

\bibitem{martinez}
M.~Martinez et~al.
\newblock These proceedings.

\bibitem{hessls}
F.~Aharonian et~al.
\newblock 3.9 day orbital modulation in the {T}e{V} gamma-ray flux and spectrum
  from the {X}-ray binary {LS} 5039.
\newblock {\em Astron.Astrophys.}, 2006.

\bibitem{1259}
F.~Aharonian et~al.
\newblock Discovery of the binary pulsar {PSR} {B}1259-63 in very-high-energy
  gamma rays around periastron with {HESS}.
\newblock {\em Astron.Astrophys.}, 2005.

\bibitem{perspectives}
I.~F. Mirabel.
\newblock Very energetic gamma-rays from microquasars and binary pulsars.
\newblock {\em Science}, 312:1759--1760, 2006.

\bibitem{gregoryold}
P.~C. Gregory and A.~R. Taylor.
\newblock {\em Nature}, 272:704--706, 1978.

\bibitem{taylor}
A.~R. Taylor et~al.
\newblock {\em Astron. Astrophys.}, 305:817--824, 1996.

\bibitem{casares}
J.~Casares, I.~Ribas, J.~M. Paredes, J.~Marti, and C.~Allende~Prieto.
\newblock Orbital parameters of the microquasar {LS I} +61 303.
\newblock {\em Mon. Not. Roy. Astron. Soc.}, 360:1091--1104, 2005.

\bibitem{gregorynew}
P.~C. Gregory.
\newblock {\em Astrophys. J.}, 575:427--434, 2002.

\bibitem{massi}
M.~Massi et~al.
\newblock Hints for a fast precessing relativistic radio jet in {LS I} +61 303.
\newblock {\em Astron. Astrophys.}, 414:L1--L4, 2004.

\bibitem{tavani}
M.~Tavani et~al.
\newblock {\em Astrophys. J.}, pages L89--L91, 1998.

\bibitem{kniffen}
D.~A. Kniffen et~al.
\newblock {\em Astrophys. J.}, pages 126--131, 1997.

\bibitem{lsi}
J.~Albert et~al.
\newblock Variable very high energy gamma-ray emission from the microquasar {LS
  I} +61 303.
\newblock {\em Science}, 312:1771--1773, 2006.

\bibitem{magic_hess1813}
J.~Albert et~al.
\newblock {MAGIC} observations of very high energy gamma-rays from {HESS
  J}1813-178.
\newblock {\em Astrophys. J.}, 637:L41--L44, 2006.

\bibitem{magic_hess1834}
J.~Albert et~al.
\newblock Observation of {VHE} gamma radiation from h{ESS} {J}1834-087/{W}41
  with the {MAGIC} telescope.
\newblock {\em Astrophys. J.}, 643:L53--L56, 2006.

\bibitem{magic_gc}
J.~Albert et~al.
\newblock Observation of gamma rays from the galactic center with the {MAGIC}
  telescope.
\newblock {\em Astrophys. J.}, 638:L101--L104, 2006.

\bibitem{rolke}
W.~A. Rolke, A.~M. Lopez, and J.~Conrad.
\newblock Confidence intervals with frequentist treatment of statistical and
  systematic uncertainties.
\newblock {\em Nucl. Instrum. Meth.}, A551:493--503, 2005.

\bibitem{frail}
D.~A. Frail and R.~M. Hjellming.
\newblock {\em Astrophys. J.}, pages 2126--2130, 1991.

\bibitem{wipple}
S.~J. Fegan et~al.
\newblock A survey of unidentified {EGRET} sources at very high energies.
\newblock {\em Astrophys. J.}, 624:638--655, 2005.

\bibitem{mirabel}
I.~F. Mirabel and L.~F. Rodriguez.
\newblock Sources of relativistic jets in the galaxy.
\newblock {\em Ann. Rev. Astron. Astrophys.}, 37:409--443, 1999.

\bibitem{heinz}
S.~Heinz and R.~A. Sunyaev.
\newblock Cosmic rays from microquasars: a narrow component to the {CR}
  spectrum?
\newblock {\em Astron. Astrophys.}, 390:751--766, 2002.

\bibitem{valenti}
V.~Bosch-Ramon, J.~M. Paredes, G.~E. Romero, and D.~F. Torres.
\newblock A microquasar model applied to unidentified gamma-ray sources.
\newblock {\em Astron. Astrophys.}, 446:1081--1088, 2006.

\bibitem{maraschi}
L.~Maraschi and A.~Treves.
\newblock {\em Mon. Not. Roy. Astron. Soc.}, page~1P, 1981.

\end{thebibliography}
\bibliographystyle{unsrt}

%


\printindex
\end{document}